# The Effects of Hofstede's Cultural Dimensions on Pro-Environmental Behaviour: How Culture Influences Environmentally Conscious Behaviour

**SZABOLCS NAGY, Ph.D.**
ASSOCIATE PROFESSOR
UNIVERSITY OF MISKOLC
e-mail: marvel@uni-miskolc.hu

**CSILLA KONYHA MOLNÁRNÉ**
PhD STUDENT
UNIVERSITY OF MISKOLC
e-mail: csilla.konyha@gmail.com

*SUMMARY*

*The need for a more sustainable lifestyle is a key focus for several countries. Using a questionnaire survey conducted in Hungary, this paper examines how culture influences environmentally conscious behaviour. Having investigated the direct impact of Hofstede's cultural dimensions on pro-environmental behaviour, we found that the culture of a country hardly affects actual environmentally conscious behaviour. The findings indicate that only individualism and power distance have a significant but weak negative impact on pro-environmental behaviour. Based on the findings, we can state that a positive change in culture is a necessary but not sufficient condition for making a country greener.*
*Journal of Economic Literature (JEL) codes: M31, P27, Q01, Q50, Z13*
*DOI: http://dx.doi.org/10.18096/TMP.2018.01.03*

## INTRODUCTION

Our world is overwhelmed with environmental and social problems. Air pollution, climate change, deforestation, extinction of species, soil degradation, chemicals and waste are regarded as the most crucial environmental issues (UNEP, 2016). Culture influences behavioural patterns of individuals, including pro-environmental behaviour, to a large extent through socialization; therefore, analysis of the effects of culture on environmentally conscious behaviour is indispensable. The starting point of our investigation is that different cultures are based on different dominant core values. Those core values determine to what extent people will behave in an environmentally conscious way and whether environmental friendly products will be accepted in a society, and if so, to what extent consumers will demand them. We assume if a culture is based on a dominant set of values that are positively correlated with pro-environmental behaviour, that is, if environmental-related values are important in a society, it will have a positive impact on the general level of pro-environmental behaviour and the demand for environmental friendly products. Our research objective was to analyse how culture influences pro-environmental behaviour.

## LITERATURE REVIEW

Although many researchers have addressed the negative consequences of individual behaviour behind environmental issues (Boldero 1995; Oskamp 2000; Nordlund & Garvill 2002; Ojala 2008; Klöckner & Oppedal 2011; Swami et al. 2011; Guerrero et al. 2013; Marshall & Farahbakhsh 2013), previous works failed to investigate the effects of Hofstede's cultural dimensions on pro-environmental behaviour. However, understanding and predicting forces influencing pro-environmental behaviour would be highly significant, as previous studies (Nagy 2005, 2012; Hofmeister-Tóth et al. 2011) suggest that the level of environmentally conscious behaviour is rather low in Hungary. Szakály et al. 2015 found that the size of environmentally conscious LOHAS (Lifestyle of Health and Sustainability) customer segment in Hungary was only 8.7 percent.

Pro-environmental behaviour is defined by Steg and Vlek (2009, p. 309) to mean "behavior that harms the environment as little as possible, or even benefits the environment". Tylor (1871) was probably the first one to define culture as "the complex whole which includes knowledge, beliefs, arts, morals, law, customs, and any other capabilities and habits acquired by [a human] as a member of society." According to Hofstede (2011) culture is "the collective programming of the mind which distinguishes the members of one group or category of people from another." Hofstede states that we can distinguish three levels in programming the mind, which are:

- universal human nature (inherited)
- group specific culture (learned)
- personality (inherited and learned)

The aim of Hofstede's early research (1980) was to globally analyse the differences in employee values. He collected data concerning culture from more than forty countries in the world, then he analysed them using statistical methods. Culture and the personality traits of the individuals are interrelated; they mutually and greatly affect each other. In the 1980s Hofstede identified four dimensions of culture as follows:

- power distance (PDI),
- uncertainty avoidance (UAI).
- individualism – collectivism (IND)
- masculinity – femininity (MAS)

Later he added a fifth dimension to his model (Hofstede & Bond 1988), which was called long term orientation (LTO), then he introduced the sixth dimension, which is indulgence – restraint (Hofstede et al. 2010). This is the development of the 6D model of national culture, which is Hofstede's latest model for exploring the similarities and differences across national cultures (Hofstede 2017). The relative positions of the countries involved in the model on these six dimensions are expressed in a score on a 0-to-100-point scale. The higher value is intended to represent the stronger presence of the given dimension in the given country.

Power distance (PDI) refers to the opinion about inequality among people and the modes of handling the problem: how much the members of a society who are excluded from power accept and expect the unequal distribution of power. In societies with high power distance not only the leaders but people who are excluded from power also support the system. In those countries the support of autocratic or oligarchic leadership is significant: power is concentrated in a narrow circle, paternalistic leadership style is expected, children are taught to obey and give respect at school and in the family. In contrast with this, low power distance societies (i.e. Scandinavian countries) show a democratic system in practice, they have pluralist governance, privileges are not accepted; children are considered to be equal in the family and at school as well (Hofstede et al. 1998). Based on the fact that Scandinavian countries are performing exceptionally well in sustainability rankings, i.e. Finland, Iceland, Sweden and Denmark are the four best performers in 2016 EPI rankings (Hsu et al. 2016), *it can be assumed that low power distance has a positive impact on environmentally conscious behaviour (Hypothesis 1).*

Uncertainty Avoidance Index (UAI) expresses the level of stress that unknown situations can cause in a society. It refers to how much people feel uncomfortable with uncertainty and ambiguity. Avoiding uncertainty is not the same as avoiding risk, since uncertainty avoidance means how a society tolerates ambiguous situations. In cultures exhibiting strong uncertainty avoidance (Latin America, Mediterranean countries and Japan) written rules, laws of behaviour are very important, the level of risk taking is low and conflict avoidance behaviour is typical. On the other hand, in cultures where the degree of uncertainty avoidance is low, uncertainty is regarded as a natural inherent of life and people consider unusual situations as opportunities rather than threats.

The individualism versus collectivism (IND) dimension of culture refers to how much individuals integrate into the primary groups, to what extent they care about only themselves and/or their close family. It expresses how responsible people feel for the members of a wider community (for example relatives), who also expect support in return. In individualist cultures (e.g. the USA, Hungary, etc.) the degree of emotional attachment to groups is low, self-reliance, diversity and self-centredness are highly important. Everyone cares for himself/herself or the immediate family. Members of collective societies (e.g. South Asia, Korea, Japan and China) are fully identified with their community from their birth. Relationships within the community are strong, cohesion is high. Loyalty toward the extended family (i.e. grandparents and relatives), which protects its members in return, is unquestionable.

Masculinity versus femininity (MAS) dimension refers to the emotional roles between men and women, as well as role-sharing of genders. In masculine societies (e.g. Hungary) masculine and feminine roles are clearly distinguished. In masculine societies "we live to work", so focusing on work and its exaggerated form, workaholism, is typical. The most important goals for people are to make achievements and to make money. The most important values in those countries are related to money and career. It is common for people to show their high status in society by owning recognised brands and luxury goods, which is in contrast with environmentally conscious behaviour. In feminine societies with modest, caring features (i.e. Scandinavian countries) protecting the environment and nature, caring for others, solidarity, the need for better quality of life and nurturing human relations are crucial (Hofstede and Arrindell 1998). *For all these reasons we suppose that masculinisation of a society is against environmentally conscious behaviour (Hypothesis 2).*

Long-term orientation versus short term normative orientation dimension (LTO) signals that the focus of human behaviour is placed on the future or present/past. In this context, it is referred to as "(short term) normative versus (long term) pragmatic" (PRA). In the academic context, the terminology Monumentalism versus

Flexhumility can also be used (Hofstede 2017). The most important distinguishing features of high level long term orientation, which is typical of China, Korea, Japan and some other Asian countries, are persistence, saving and shaming those who do not fulfil duties. People with such an attitude think that the most important events of life have not happened yet, they will occur in the future. The ability to change is important for them. It means that a "good" person adapts to the circumstances. This is true for traditions as well. Traditions must be adjusted to the circumstances. Such cultures are characterised by learning from others. In contrast, in western societies with short term orientation, people tend to think that the most important things are happening now or have already happened. Such cultures have sacred and inviolable traditions. People are proud of their own nation and do not want to change their traditions. Learning from others is not typical for them (Hofstede and Bond 1988).

The indulgence (IND) versus restraint dimension focuses on how people satisfy or control the basic human drive for an enjoyable life. In societies exhibiting strong indulgence people are allowed to freely satisfy their desires in connection with enjoying life and having fun. On the other hand, in societies where the level of restraint is high, strict social norms regulate the gratification of needs. In restrained societies only a few people are happy; many of them feel they are vulnerable because things just happen to them. Spare time and comfort are not priorities in restrained countries. Fewer people do sports, sexual norms are stricter, the birth rate is lower but there are also fewer obese people than in cultures permitting an enjoyable life (Hofstede et al. 2010).

Onel and Mukherjee (2013) investigated the effects of national culture and human development on environmental health. Using multiple linear regression models, they found that cultural dimensions of individualism and uncertainty avoidance, as well as human development components of life expectancy at birth, education, and income, significantly influence environmental health performance.

Cho et al. (2013) investigated the relationship between collectivism versus individualism as a cultural dimension and environmentally conscious behaviour by using the value-belief-norm model. They found that both horizontal collectivism – when the individual is the part of the group and there are no differences among the individuals within the group - and vertical individualism – when the individual is autonomous, independent and accepts differences - are important influential factors of perceived consumer effectiveness, which has a positive effect on environmental attitudes and finally results in higher levels of environmentally conscious commitment.

Once and Almagtome (2014) made a cross-cultural comparison of the effect of national culture values on corporate environmental disclosure (CED). They found that two of Hofstede's national culture dimensions were linked to a higher degree of corporate environmental disclosure. In particular, a nation's high degree of individualism and long-term orientation were linked to high levels of corporate environmental disclosure. On the other hand, they found that one of Hofstede's national culture dimensions were related to a low degree of corporate environmental disclosure.

## DATA AND METHODS

In order to investigate pro-environmental behaviour an online survey was conducted in Hungary in 2017. A total of 442 respondents aged over 18 were included in the convenience sample with the snowball method. This means a 4.66% confidence interval at the 95% confidence level. As the original sample was not a representative sample, we used a commonly applied correction technique, the weighting adjustment, to make our sample representative according to variables such as sex and age.

To explore the impact of culture on pro-environmental behaviour, we investigated the relationships between Hofstede's cultural dimensions (HCD) and environmentally conscious behaviour. In an attempt to measure pro-environmental behaviour, we used a revised version of the General Ecological Behaviour scale. The original measuring tool involves thirty-eight items in two sections representing different types of ecological and pro-social behaviour (Kaiser et al. 1999). Since we did not intend to investigate pro-social behaviour and some of the pro-environmental items proved to be irrelevant or outdated (Nagy, 2012), we deliberately left out eight variables concerning prosocial behaviour and three variables regarding ecological behaviour from the revised version of the GEB scale. However, we added ten ecological behaviour items, therefore the resulting pro-environmental behaviour scale (PEB scale) consists of thirty-seven items (Appendix 1).

We measured the actual behaviour instead of behaviour intention by using dichotomous questions (yes/no responses). Negative behaviour items (item No: 5, 7, 10, 11, 12, 13, 16, 19, 21, 22, 23, 30, 35 and 36) were reversed in coding. Missing values were handled as 'no' responses. Behaviour difficulty of each PEB item was calculated by dividing the number of people behaving in an environmentally conscious way by the total number of respondents. We also considered the respondents' tendency to behave ecologically by considering the number of ecological behaviours they have carried out. In order to measure pro-environmental behaviour of individuals, we calculated the weighted sum of each item on the revised GEB scale. Difficulty parameters of pro-environmental behaviour items were used as weights. Then we divided the weighted sums by the total sum of difficulty parameters to transform it into a 0-1 scale of pro-environmental behaviour. Zero (0) score expresses that the individual does not behave environmentally consciously at all. On the contrary, if someone's behaviour is a hundred percent environmentally conscious, the PEB score will be the maximum (1).

Szabolcs Nagy – Csilla Konyha Molnárné*Table 1*
*The relationship between Hofstede's cultural dimensions and pro- environmental behaviour*

| Cultural Dimension | Operationalization | Short Form | Means ($\bar{x}$) (1-5 scale) | Pearson correlation (r) | Relationship with Pro Environmental Behaviour |
|---|---|---|---|---|---|
| **Masculinity Versus Femininity** | Competition, success and performance are more important than caring for others and the quality of life. | MAS | 1.67 | -0.084 | not significant |
| **Uncertainty Avoidance Index** | Change and unknown situations always involve a lot more threats than opportunities. | UAI | 2.31 | 0.020 | not significant |
| **Power Distance Index** | Power is distributed unequally in the society, i.e. there are the rich and the poor, and it is completely acceptable for me. | PDI | 2.43 | -.156** | weak, negative |
| **Indulgence Versus Restraint** | Social norms and expectations must always be met and we must control our desires, even if it makes our life less enjoyable. | IND | 2.55 | -0.062 | not significant |
| **Individualism Versus Collectivism** | Everyone has to take care of themselves, we cannot expect help from others. | IDV | 2.73 | -.163** | weak, negative |
| **Long Term Orientation Versus Short Term Normative Orientation** | We must rely on the past only to the extent that it serves the interests of the future. | LTO | 3.24 | -0.069 | not significant |

Notes: ** Correlation is significant at 0.01 level (2-tale)
Source: Authors' own research

Since the multidimensional measuring scale of Hofstede's cultural dimensions was not available to us, we used a simplified, one-dimensional measurement approach, therefore we measured each cultural dimension with only one variable. We used Likert's five-level scale to measure Hofstede's cultural dimensions. We asked respondents to express to what extent they agree with the statements that can be seen in the "operationalization" column in Table 1. The lowest score (1) on the Likert scale signals that the respondent did not agree with the given statement at all, while the highest score (5) indicates that (s)he completely agreed.

Then we transformed the scores that we measured on the Likert scale to a 0-100 scale to make them comparable with Hofstede's scores as the relative positions of the countries on all cultural dimensions are expressed in a score on a 0-to-100-point scale in Hofstede's 6D model. Scores below 50 points indicate the dominance of one of the values, whereas scores above 50 points refer to the dominance of the opposite value. Uncertainty resulting from this measurement transformation can be considered as a limitation of our study. Application of a scale between 0 and 100 points to measure Hofstede's cultural dimensions in future research can reduce this kind of measurement error. Because of the limitations discussed above, our results require further investigations in the future.

## RESULTS & DISCUSSION

As a result of our investigations on pro-environmental behaviour, we found that most Hungarians do not behave environmentally consciously at all, i.e. they do not consider the environmental consequences of their behaviour. Figure 1 shows the percentage distribution of the population in Hungary in terms of the level of pro-environmental behaviour. Axis x shows the level of PEB, while we can see the percentage distribution of respondents on axis y. The mean of PEB in Hungary is only 0.445 on the 0-1 PEB scale, therefore it can be concluded that the level of pro-environmental behaviour is moderately low. Our result confirms previous findings in the literature (Hofmeister-Tóth et al. 2011; Nagy 2005, 2012) that pro-environmental behaviour is not typical of Hungarians.

*30*

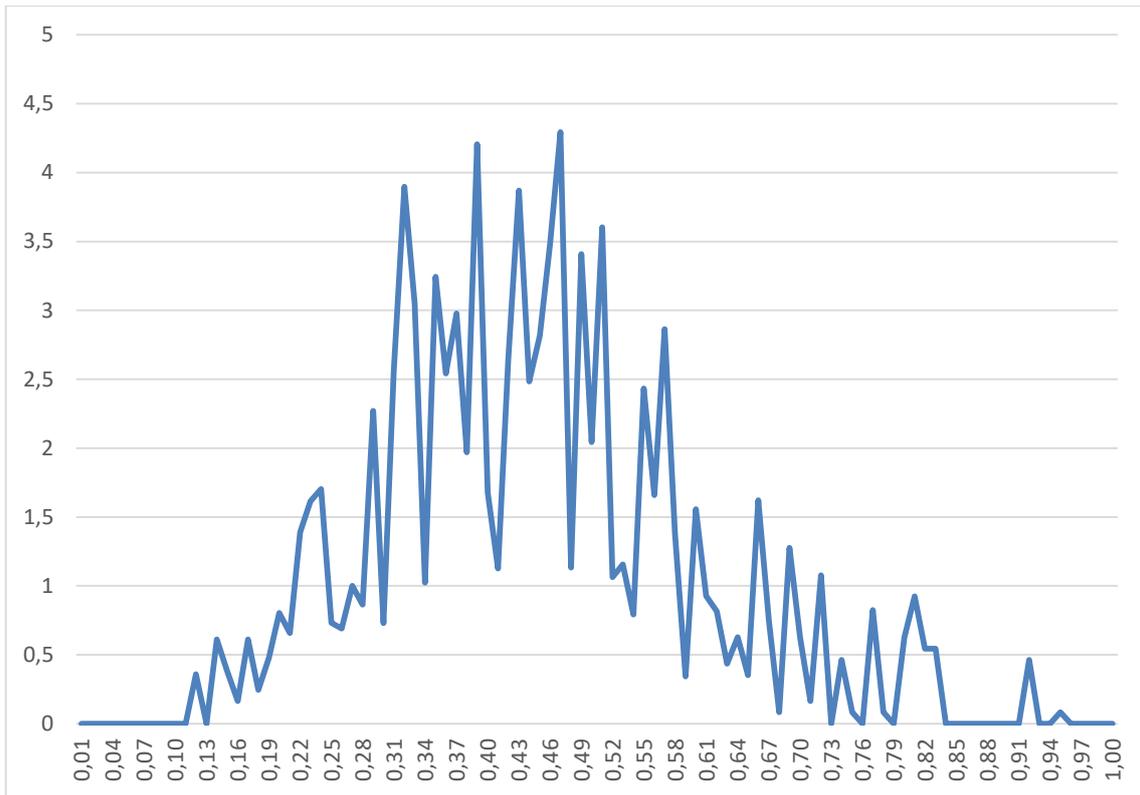

Source: Authors' own research

*Figure 1. Percentage distribution of pro-environmental behaviour (PEB) in Hungary*

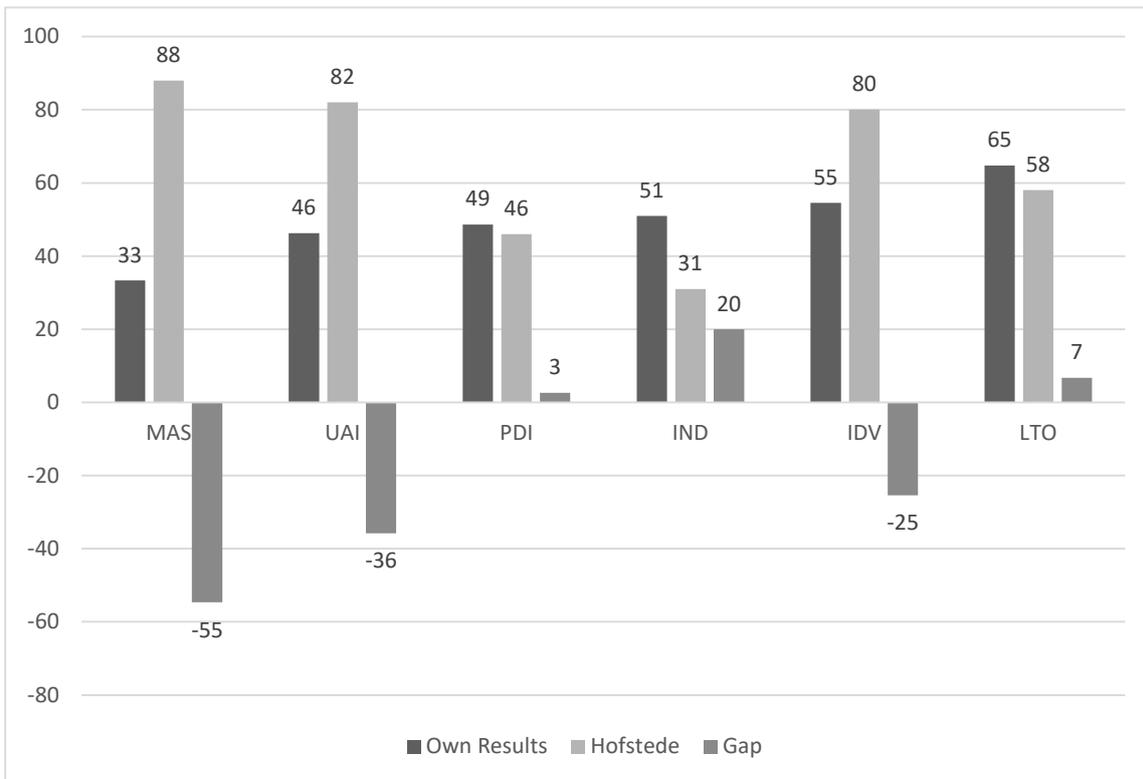

Source: Authors' own research based on Hofstede (2017)

*Figure 2. Hofstede's scores, our results and the gap
between them for Hofstede's cultural dimensions in Hungary*

Figure 2 highlights the gap between the country scores of Hungary in Hofstede's 6D model and our results. Hofstede's scores suggest that Hungary is an extremely masculine country (MAS=88), characterized by a high degree of uncertainty avoidance (UAI=82) and high level of individualism (IDV=80). The relatively high score of long term orientation (LTO=58) suggests that the Hungarian society is rather pragmatic. Power distance index (PDI=46) is relatively low, which indicates that people in Hungary are slightly against the unequal distribution of the power. Since the value of indulgence (IND=31) is very low, we can conclude that restraint is characteristic of Hungarians to a large extent. It means that many of them may think that fulfilling their desires is against social norms and an enjoyable life is "something wrong".

The most important gap – 55 points – between our results and Hofstede's scores was found in terms of the masculinity-femininity dimension, where our score (MAS=33) was significantly lower than Hofstede's score (MAS=88). Our result suggests that Hungary is a feminine country where taking care of others and quality of life are dominant values. In feminine societies a focus on quality of life is top priority and only very few people want to stand out from the crowd. While in masculine countries people are driven to be the best, in feminine cultures it is important for people to like what they do and find it interesting.

The second greatest gap (36 points) was found in terms of the uncertainty avoidance index. We measured only 46 points in contrast to Hofstede's 82 points. Another study by Neumann-Bódi et al. (2008) yielded a score of 64 points. When the UAI score is below 50 points, people are not afraid of changes and these are seen as opportunities rather than threats. In countries with low uncertainty avoidance, people feel they can shape the future to some extent and it does not just happen to them. In societies accepting uncertainty there is a willingness to accept new ideas, to try new products and entrepreneurial spirit of people is also higher. These cultures require fewer rules and people show their emotions less expressively.

The third largest gap (25 points) occurred in terms of individualism (IDV). We measured only 55 points instead of the 80 points that can be found in the 6D model. It means that Hungarian society is less individualist according to our results. Our findings are not in line with those of Neumann-Bódi et al. (2008), who found a very high level of individualism in Hungary.

However, it must be highlighted that our result is consistent with that of Hofstede's in regard to the finding that the Hungarian society is not collectivist. In Hungary, people take care of their immediate family and only loose social ties exist. Self-centredness is also characteristic in individualist societies. People need a private sphere and relationships are based on obtaining mutual benefits.

As far as indulgence and restraint are concerned, we measured much higher scores (51 points) than Hofstede did (31 points). It means that Hungarians are not so restrained as it would appear based on Hofstede's scores. We enjoy our life much more and we live it more impulsively and people do not tend to be so cynical and pessimistic.

As for the other dimensions, no significant differences can be found between our results and those of Hofstede's. The Long Term Orientation score that we measured (65 points) was only slightly higher than Hofstede's 58 points. The above results suggest that Hungary is a pragmatic country where people are convinced that truth largely depends on the specific situation, context and time. In Hungary traditions are transformed according to the changing situations and people fight persistently to achieve results.

As for Power Distance Index (PDI), the difference was insignificant, only 3 points, as we measured 49 points instead of 46 points that can be found in the 6D model. Moderately low scores for power distance mean that Hungarians tend to favour independence and do not like subordination and control. The majority of Hungarians believe in equal rights.

To test our hypotheses and to investigate the impact of Hofstede's cultural dimensions (HCD) on pro-environmental behaviour (PEB), we used the Pearson correlation. The results of the significance analysis as well as the correlation coefficients suggest that only two cultural dimensions have a significant impact on environmentally conscious behaviour. However, in both cases the relationship is only weak. *A higher level of individualism, i.e. the individualisation of the society, is slightly against pro-environmental behaviour (r=-0.163) and so is the higher level of power distance (r=-0.156).* We also found that the other cultural dimensions in Hofstede's 6D model have no significant effect on environmentally conscious behaviour (Table 1).

The results of stepwise linear regression support that *Hofstede's cultural dimensions have only a very weak direct influence on pro-environmental behaviour.* It can be assumed that HCD affects PEB indirectly through other factors (i.e. personal values and attitudes) as the regression model has only very low explanatory power (Table 2). With model 2, only 3.4 percent of the variation in the dependent variable (pro-environmental behaviour) can be explained using the independent variables (Individualism and Power Distance Index).

*Table 2*
*Regression model*

**Model Summary**

| Model | R | R Square | Adjusted R Square | Std. Error of the Estimate |
|---|---|---|---|---|
| 1 | .167[a] | .028 | .026 | .14516 |
| 2 | .197[b] | .039 | .034 | .14452 |

a. Predictors: (Constant), IDV
b. Predictors: (Constant), IDV, PDI
Source: Authors' own research

*Table 3*
*ANOVA table*

**ANOVA[a]**

| Model | | Sum of Squares | df | Mean Square | F | Significance |
|---|---|---|---|---|---|---|
| 1 | Regression | .269 | 1 | .269 | 12.749 | .000[b] |
|   | Residual | 9.313 | 442 | .021 | | |
|   | Total | 9.582 | 443 | | | |
| 2 | Regression | .371 | 2 | .185 | 8.881 | .000[c] |
|   | Residual | 9.211 | 441 | .021 | | |
|   | Total | 9.582 | 443 | | | |

a. Dependent Variable: PEB (0-1)
b. Predictors: (Constant), IDV
c. Predictors: (Constant), IDV, PDI
Source: Authors' own research

*Table 4*
*Regression - Coefficients*

**Coefficients[a]**

| Model | | Unstandardized Coefficients | | Standardized Coefficients | t | Sig. |
|---|---|---|---|---|---|---|
| | | B | Std. Error | Beta | | |
| 1 | (Constant) | .495 | .018 | | 27.412 | .000 |
|   | IDV | -.021 | .006 | -.167 | -3.571 | .000 |
| 2 | (Constant) | .513 | .020 | | 26.042 | .000 |
|   | IDV | -.016 | .006 | -.126 | -2.513 | .012 |
|   | PDI | -.013 | .006 | -.111 | -2.214 | .027 |

a. Dependent Variable: PEB (0-1)
Source: Authors' own research

Significance values in Table 3 indicate that both models are significant.

As Table 4 shows, individualism has a weak negative effect on pro-environmental behaviour (β=-0.126), while higher Power Distance is also against pro-environmental behaviour (β=-0.111). B scores in Table 4 suggest that respondents who score 1 point higher on Individualism will – on average – score 0.16 points lower on the PEB scale, while people who score 1 point higher on Power Distance Index will - on average - score 0.13 points lower on the PEB scale.

Based on the above findings, it can be concluded that *in collectivist societies with low power distance the probability of pro-environmental behaviour is higher.* Both the results of Pearson correlation and the linear regression support our first hypothesis, as we found that *high power distance index has a negative impact on pro-environmental behaviour.* However, the second hypothesis is not supported, since we found no significant relationship between the feminine/masculine nature of a society and the level of PEB.

## CONCLUSIONS

This research was carried out in order to analyse how culture influences pro-environmental behaviour. Firstly,

we investigated the level of environmental consciousness in Hungary. We measured actual behaviour instead of behaviour intention and found that the level of pro-environmental behaviour is moderately low. This means that corrective actions are needed to increase environmentally consciousness. However, changing the culture of the country would not be sufficient, as the evidence from this study suggests that Hofstede's cultural dimensions only slightly influence pro-environmental behaviour. Among the significant cultural dimensions, only individualism and power distance have a weak negative impact on environmentally conscious behaviour. Yet, for these reasons, if we intend to make a country greener, collectivization of the society – or at least significant moderation of the level of individualism – and/or lowering the level of power distance are required. The results also suggest that people living in countries with low individualism and power distance index (e.g. Costa Rica) will behave in a more environmentally conscious way. The 2014 Global Green Economy Index also confirms this interpretation, as Costa Rica recorded an impressive result, ranking 3rd behind Sweden and Norway on performance and in the top 15 for perceptions overall (GGEI, 2014).

# ACKNOWLEDGMENT

"The described article was carried out as part of the EFOP-3.6.1- 16-2016-00011 "Younger and Renewing University – Innovative Knowledge City – institutional development of the University of Miskolc aiming at intelligent specialisation" project implemented in the framework of the Szechenyi 2020 program. The realization of this project is supported by the European Union, co-financed by the European Social Fund."

"A cikkben ismertetett kutató munka az EFOP-3.6.1-16-2016-00011 jelű „Fiatalodó és Megújuló Egyetem – Innovatív Tudásváros – a Miskolci Egyetem intelligens szakosodást szolgáló intézményi fejlesztése" projekt részeként – a Széchenyi 2020 keretében – az Európai Unió támogatásával, az Európai Szociális Alap társfinanszírozásával valósul meg"

*Appendix 1*

Pro-environmental scale items

1. After meals, I dispose of leftovers in the toilet.*
2. For shopping, I prefer paper bags to plastic ones.
3. I am a member of an environmental organization.
4. I bring empty bottles to a recycling bin.
5. I buy a lot of products made of recycled materials.
6. I collect and recycle used paper.
7. I do not buy anything from companies being not socially or environmentally responsible.
8. I do not buy products tested on animals.
9. I do not change anything just because it is out of fashion.
10. I often talk with friends about problems related to the environment.
11. I prefer local products and foods to those transported from faraway areas.
12. I prefer to shower rather than to take a bath.
13. I put dead batteries in the garbage.*
14. I put unused medicine in the dustbin.
15. I sometimes contribute financially to environmental organizations.
16. I travel by air at least once or twice a year.*
17. I use a chemical air freshener in my bathroom.*
18. I use alternatives (e.g. washing nuts) for washing instead of detergents.
19. I use an oven-cleaning spray to clean my oven.*
20. I use chemical toilet-cleaners.*
21. I use fabric softener with my laundry.*
22. I usually buy environmentally-friendly products or organic foods.
23. I usually buy milk in returnable bottles.
24. I usually drive on motorways at speeds under 100 km/h.
25. I wait until I have a full load before doing my laundry.
26. I wash dirty clothes without prewashing.
27. I eat meat at every meal*
28. If I am offered a plastic bag in a store I will always take it.*
29. If there are insects in my apartment I kill them with a chemical insecticide.*
30. In the past, I have pointed out to someone his or her unecological behaviour.
31. In the winter, I keep the heat on so that I do not have to wear a sweater.*
32. In the winter, I leave the windows open for long periods of time to let in fresh air.*
33. Sometimes I buy beverages in cans.*
34. There is significantly less waste in my household than a year ago.
35. Usually I do not drive my automobile in the city.
36. When buying a new household device, I always prefer the more energy efficient versions.
37. When possible within short distances, I use public transportation or ride a bike.

* Negative behaviour items.